\documentclass[conference,a4paper]{IEEEtran}

\usepackage[english]{babel}
\usepackage[utf8]{inputenc}
\usepackage[T1]{fontenc}
\usepackage{hyphenat}

\usepackage{times}
\usepackage{amsmath,amssymb,amstext,amsfonts,mathrsfs,amsthm,mathtools,cases}
\usepackage{comment}

\usepackage{empheq}

\usepackage{graphicx}                           
\graphicspath{{.}{./Figures/}}

\usepackage[nolist,printonlyused]{acronym}      

\usepackage[sort]{cite}                         

\usepackage{enumerate}                          

\usepackage{multirow}
\usepackage{tabularx}
\usepackage{booktabs}
\newcolumntype{C}{>{\centering\arraybackslash}X}
\newcolumntype{R}{>{\flushright\arraybackslash}X}
\newcolumntype{L}{>{\flushleft\arraybackslash}X}

\usepackage{algorithmic}
\usepackage[ruled]{algorithm}
\algsetup{indent=2ex}
\usepackage{eqparbox}

\usepackage{setspace}
\usepackage{xspace} 
\newcommand*{\unit}[2]{\mbox{\ensuremath{#1\,\mathrm{#2}}}}

\usepackage{float}
\floatname{algorithm}{\footnotesize Algorithm}

\usepackage{url}

\usepackage[normalem]{ulem}

\newcommand*{\figref}[1]{Figure~\ref{#1}}
\newcommand*{\secref}[1]{Section~\ref{#1}}
\newcommand*{\tabref}[1]{Table~\ref{#1}}
\newcommand*{\algref}[1]{Algorithm~\ref{#1}}

\theoremstyle{definition}

\DeclareMathOperator*{\argmax}{arg\,max}

\usepackage{color}

\newcommand{\comm}[1]{}

\IEEEoverridecommandlockouts

\allowdisplaybreaks[4]

\begin{document}
\date{\today}

\bstctlcite{IEEEexample:BSTcontrol}

\title{On the Spectral Efficiency and Fairness in Full-Duplex Cellular Networks}

\author{
Jos\'{e} Mairton B.~da Silva Jr$^{\star}$
\thanks{Jos\'{e} Mairton B. da Silva Jr. was supported by 
CNPq, a Brazilian research-support
agency, and by The Lars Hierta Memorial Foundation.
The simulations were performed on resources provided by the Swedish National Infrastructure for
Computing (SNIC) at PDC Centre for High Performance Computing (PDC-HPC).
G\'{a}bor Fodor was supported by the Wireless@KTH Project BUSE.}\hspace{-2mm},
G\'{a}bor Fodor$^{\star\dagger}$,
Carlo Fischione${^\star}$\\
$^\star$KTH and Electrical Engineering School, Royal Institute of Technology, Stockholm, Sweden \\
$^\dagger$Ericsson Research, Kista, Sweden
\\[0pt]\vspace{-0.95\baselineskip}
}


\begin{acronym}[LTE-Advanced]
  \acro{2G}{Second Generation}
  \acro{3-DAP}{3-Dimensional Assignment Problem}
  \acro{3G}{3$^\text{rd}$~Generation}
  \acro{3GPP}{3$^\text{rd}$~Generation Partnership Project}
  \acro{4G}{4$^\text{th}$~Generation}
  \acro{5G}{5$^\text{th}$~Generation}
  \acro{AA}{Antenna Array}
  \acro{AC}{Admission Control}
  \acro{AD}{Attack-Decay}
  \acro{ADSL}{Asymmetric Digital Subscriber Line}
  \acro{AHW}{Alternate Hop-and-Wait}
  \acro{AMC}{Adaptive Modulation and Coding}
  \acro{AP}{Access Point}
  \acro{APA}{Adaptive Power Allocation}
  \acro{ARMA}{Autoregressive Moving Average}
  \acro{ATES}{Adaptive Throughput-based Efficiency-Satisfaction Trade-Off}
  \acro{AWGN}{Additive White Gaussian Noise}
  \acro{BB}{Branch and Bound}
  \acro{BD}{Block Diagonalization}
  \acro{BER}{Bit Error Rate}
  \acro{BF}{Best Fit}
  \acro{BFD}{bidirectional full duplex}
  \acro{BLER}{BLock Error Rate}
  \acro{BPC}{Binary Power Control}
  \acro{BPSK}{Binary Phase-Shift Keying}
  \acro{BRA}{Balanced Random Allocation}
  \acro{BS}{base station}
  \acro{CAP}{Combinatorial Allocation Problem}
  \acro{CAPEX}{Capital Expenditure}
  \acro{CBF}{Coordinated Beamforming}
  \acro{CBR}{Constant Bit Rate}
  \acro{CBS}{Class Based Scheduling}
  \acro{CC}{Congestion Control}
  \acro{CDF}{Cumulative Distribution Function}
  \acro{CDMA}{Code-Division Multiple Access}
  \acro{CL}{Closed Loop}
  \acro{CLPC}{Closed Loop Power Control}
  \acro{CNR}{Channel-to-Noise Ratio}
  \acro{CPA}{Cellular Protection Algorithm}
  \acro{CPICH}{Common Pilot Channel}
  \acro{CoMP}{Coordinated Multi-Point}
  \acro{CQI}{Channel Quality Indicator}
  \acro{CRM}{Constrained Rate Maximization}
	\acro{CRN}{Cognitive Radio Network}
  \acro{CS}{Coordinated Scheduling}
  \acro{CSI}{Channel State Information}
  \acro{CUE}{Cellular User Equipment}
  \acro{D2D}{device-to-device}
  \acro{DCA}{Dynamic Channel Allocation}
  \acro{DE}{Differential Evolution}
  \acro{DFT}{Discrete Fourier Transform}
  \acro{DIST}{Distance}
  \acro{DL}{downlink}
  \acro{DMA}{Double Moving Average}
	\acro{DMRS}{Demodulation Reference Signal}
  \acro{D2DM}{D2D Mode}
  \acro{DMS}{D2D Mode Selection}
  \acro{DPC}{Dirty Paper Coding}
  \acro{DRA}{Dynamic Resource Assignment}
  \acro{DSA}{Dynamic Spectrum Access}
  \acro{DSM}{Delay-based Satisfaction Maximization}
  \acro{ECC}{Electronic Communications Committee}
  \acro{EFLC}{Error Feedback Based Load Control}
  \acro{EI}{Efficiency Indicator}
  \acro{eNB}{Evolved Node B}
  \acro{EPA}{Equal Power Allocation}
  \acro{EPC}{Evolved Packet Core}
  \acro{EPS}{Evolved Packet System}
  \acro{E-UTRAN}{Evolved Universal Terrestrial Radio Access Network}
  \acro{ES}{Exhaustive Search}
  \acro{FD}{full-duplex}
  \acro{FDD}{frequency division duplex}
  \acro{FDM}{Frequency Division Multiplexing}
  \acro{FER}{Frame Erasure Rate}
  \acro{FF}{Fast Fading}
  \acro{FSB}{Fixed Switched Beamforming}
  \acro{FST}{Fixed SNR Target}
  \acro{FTP}{File Transfer Protocol}
  \acro{GA}{Genetic Algorithm}
  \acro{GBR}{Guaranteed Bit Rate}
  \acro{GLR}{Gain to Leakage Ratio}
  \acro{GOS}{Generated Orthogonal Sequence}
  \acro{GPL}{GNU General Public License}
  \acro{GRP}{Grouping}
  \acro{HARQ}{Hybrid Automatic Repeat Request}
  \acro{HD}{half-duplex}
  \acro{HMS}{Harmonic Mode Selection}
  \acro{HOL}{Head Of Line}
  \acro{HSDPA}{High-Speed Downlink Packet Access}
  \acro{HSPA}{High Speed Packet Access}
  \acro{HTTP}{HyperText Transfer Protocol}
  \acro{ICMP}{Internet Control Message Protocol}
  \acro{ICI}{Intercell Interference}
  \acro{ID}{Identification}
  \acro{IETF}{Internet Engineering Task Force}
  \acro{ILP}{Integer Linear Program}
  \acro{JRAPAP}{Joint RB Assignment and Power Allocation Problem}
  \acro{UID}{Unique Identification}
  \acro{IID}{Independent and Identically Distributed}
  \acro{IIR}{Infinite Impulse Response}
  \acro{ILP}{Integer Linear Problem}
  \acro{IMT}{International Mobile Telecommunications}
  \acro{INV}{Inverted Norm-based Grouping}
	\acro{IoT}{Internet of Things}
  \acro{IP}{Integer Programming}
  \acro{IPv6}{Internet Protocol Version 6}
  \acro{ISD}{Inter-Site Distance}
  \acro{ISI}{Inter Symbol Interference}
  \acro{ITU}{International Telecommunication Union}
  \acro{JAFM}{joint assignment and fairness maximization}
  \acro{JAFMA}{joint assignment and fairness maximization algorithm}
  \acro{JOAS}{Joint Opportunistic Assignment and Scheduling}
  \acro{JOS}{Joint Opportunistic Scheduling}
  \acro{JP}{Joint Processing}
	\acro{JS}{Jump-Stay}
  \acro{KKT}{Karush-Kuhn-Tucker}
  \acro{L3}{Layer-3}
  \acro{LAC}{Link Admission Control}
  \acro{LA}{Link Adaptation}
  \acro{LC}{Load Control}
  \acro{LOS}{line of sight}
  \acro{LP}{Linear Programming}
  \acro{LTE}{Long Term Evolution}
	\acro{LTE-A}{\ac{LTE}-Advanced}
  \acro{LTE-Advanced}{Long Term Evolution Advanced}
  \acro{M2M}{Machine-to-Machine}
  \acro{MAC}{medium access control}
  \acro{MANET}{Mobile Ad hoc Network}
  \acro{MC}{Modular Clock}
  \acro{MCS}{Modulation and Coding Scheme}
  \acro{MDB}{Measured Delay Based}
  \acro{MDI}{Minimum D2D Interference}
  \acro{MF}{Matched Filter}
  \acro{MG}{Maximum Gain}
  \acro{MH}{Multi-Hop}
  \acro{MIMO}{Multiple Input Multiple Output}
  \acro{MINLP}{mixed integer nonlinear programming}
  \acro{MIP}{Mixed Integer Programming}
  \acro{MISO}{Multiple Input Single Output}
  \acro{MLWDF}{Modified Largest Weighted Delay First}
  \acro{MME}{Mobility Management Entity}
  \acro{MMSE}{Minimum Mean Square Error}
  \acro{MOS}{Mean Opinion Score}
  \acro{MPF}{Multicarrier Proportional Fair}
  \acro{MRA}{Maximum Rate Allocation}
  \acro{MR}{Maximum Rate}
  \acro{MRC}{Maximum Ratio Combining}
  \acro{MRT}{Maximum Ratio Transmission}
  \acro{MRUS}{Maximum Rate with User Satisfaction}
  \acro{MS}{Mode Selection}
  \acro{MSE}{Mean Squared Error}
  \acro{MSI}{Multi-Stream Interference}
  \acro{MTC}{Machine-Type Communication}
  \acro{MTSI}{Multimedia Telephony Services over IMS}
  \acro{MTSM}{Modified Throughput-based Satisfaction Maximization}
  \acro{MU-MIMO}{Multi-User Multiple Input Multiple Output}
  \acro{MU}{Multi-User}
  \acro{NAS}{Non-Access Stratum}
  \acro{NB}{Node B}
	\acro{NCL}{Neighbor Cell List}
  \acro{NLP}{Nonlinear Programming}
  \acro{NLOS}{non-line of sight}
  \acro{NMSE}{Normalized Mean Square Error}
  \acro{NORM}{Normalized Projection-based Grouping}
  \acro{NP}{non-polynomial time}
  \acro{NRT}{Non-Real Time}
  \acro{NSPS}{National Security and Public Safety Services}
  \acro{O2I}{Outdoor to Indoor}
  \acro{OFDMA}{Orthogonal Frequency Division Multiple Access}
  \acro{OFDM}{Orthogonal Frequency Division Multiplexing}
  \acro{OFPC}{Open Loop with Fractional Path Loss Compensation}
	\acro{O2I}{Outdoor-to-Indoor}
  \acro{OL}{Open Loop}
  \acro{OLPC}{Open-Loop Power Control}
  \acro{OL-PC}{Open-Loop Power Control}
  \acro{OPEX}{Operational Expenditure}
  \acro{ORB}{Orthogonal Random Beamforming}
  \acro{JO-PF}{Joint Opportunistic Proportional Fair}
  \acro{OSI}{Open Systems Interconnection}
  \acro{PAIR}{D2D Pair Gain-based Grouping}
  \acro{PAPR}{Peak-to-Average Power Ratio}
  \acro{P2P}{Peer-to-Peer}
  \acro{PC}{Power Control}
  \acro{PCI}{Physical Cell ID}
  \acro{PDCCH}{physical downlink control channel}
  \acro{PDF}{Probability Density Function}
  \acro{PER}{Packet Error Rate}
  \acro{PF}{Proportional Fair}
  \acro{P-GW}{Packet Data Network Gateway}
  \acro{PL}{Pathloss}
  \acro{PRB}{Physical Resource Block}
  \acro{PROJ}{Projection-based Grouping}
  \acro{ProSe}{Proximity Services}
  \acro{PS}{Packet Scheduling}
  \acro{PSO}{Particle Swarm Optimization}
  \acro{PUCCH}{physical uplink control channel}
  \acro{PZF}{Projected Zero-Forcing}
  \acro{QAM}{Quadrature Amplitude Modulation}
  \acro{QoS}{quality of service}
  \acro{QPSK}{Quadri-Phase Shift Keying}
  \acro{RAISES}{Reallocation-based Assignment for Improved Spectral Efficiency and Satisfaction}
  \acro{RAN}{Radio Access Network}
  \acro{RA}{Resource Allocation}
  \acro{RAT}{Radio Access Technology}
  \acro{RATE}{Rate-based}
  \acro{RB}{resource block}
  \acro{RBG}{Resource Block Group}
  \acro{REF}{Reference Grouping}
  \acro{RF}{Radio-Frequency}
  \acro{RLC}{Radio Link Control}
  \acro{RM}{Rate Maximization}
  \acro{RNC}{Radio Network Controller}
  \acro{RND}{Random Grouping}
  \acro{RRA}{Radio Resource Allocation}
  \acro{RRM}{Radio Resource Management}
  \acro{RSCP}{Received Signal Code Power}
  \acro{RSRP}{Reference Signal Receive Power}
  \acro{RSRQ}{Reference Signal Receive Quality}
  \acro{RR}{Round Robin}
  \acro{RRC}{Radio Resource Control}
  \acro{RSSI}{Received Signal Strength Indicator}
  \acro{RT}{Real Time}
  \acro{RU}{Resource Unit}
  \acro{RUNE}{RUdimentary Network Emulator}
  \acro{RV}{Random Variable}
  \acro{SAC}{Session Admission Control}
  \acro{SCM}{Spatial Channel Model}
  \acro{SC-FDMA}{Single Carrier - Frequency Division Multiple Access}
  \acro{SD}{Soft Dropping}
  \acro{S-D}{Source-Destination}
  \acro{SDPC}{Soft Dropping Power Control}
  \acro{SDMA}{Space-Division Multiple Access}
  \acro{SER}{Symbol Error Rate}
  \acro{SES}{Simple Exponential Smoothing}
  \acro{S-GW}{Serving Gateway}
  \acro{SINR}{signal-to-interference-plus-noise ratio}
  \acro{SI}{self-interference}
  \acro{SIP}{Session Initiation Protocol}
  \acro{SISO}{Single Input Single Output}
  \acro{SIMO}{Single Input Multiple Output}
  \acro{SIR}{Signal to Interference Ratio}
  \acro{SLNR}{Signal-to-Leakage-plus-Noise Ratio}
  \acro{SMA}{Simple Moving Average}
  \acro{SNR}{Signal to Noise Ratio}
  \acro{SORA}{Satisfaction Oriented Resource Allocation}
  \acro{SORA-NRT}{Satisfaction-Oriented Resource Allocation for Non-Real Time Services}
  \acro{SORA-RT}{Satisfaction-Oriented Resource Allocation for Real Time Services}
  \acro{SPF}{Single-Carrier Proportional Fair}
  \acro{SRA}{Sequential Removal Algorithm}
  \acro{SRS}{Sounding Reference Signal}
  \acro{SU-MIMO}{Single-User Multiple Input Multiple Output}
  \acro{SU}{Single-User}
  \acro{SVD}{Singular Value Decomposition}
  \acro{TCP}{Transmission Control Protocol}
  \acro{TDD}{time division duplex}
  \acro{TDMA}{Time Division Multiple Access}
  \acro{TNFD}{three node full-duplex}
  \acro{TETRA}{Terrestrial Trunked Radio}
  \acro{TP}{Transmit Power}
  \acro{TPC}{Transmit Power Control}
  \acro{TTI}{Transmission Time Interval}
  \acro{TTR}{Time-To-Rendezvous}
  \acro{TSM}{Throughput-based Satisfaction Maximization}
  \acro{TU}{Typical Urban}
  \acro{UE}{user equipment}
  \acro{UEPS}{Urgency and Efficiency-based Packet Scheduling}
  \acro{UL}{uplink}
  \acro{UMTS}{Universal Mobile Telecommunications System}
  \acro{URI}{Uniform Resource Identifier}
  \acro{URM}{Unconstrained Rate Maximization}
  \acro{VR}{Virtual Resource}
  \acro{VoIP}{Voice over IP}
  \acro{WAN}{Wireless Access Network}
  \acro{WCDMA}{Wideband Code Division Multiple Access}
  \acro{WF}{Water-filling}
  \acro{WiMAX}{Worldwide Interoperability for Microwave Access}
  \acro{WINNER}{Wireless World Initiative New Radio}
  \acro{WLAN}{Wireless Local Area Network}
  \acro{WMPF}{Weighted Multicarrier Proportional Fair}
  \acro{WPF}{Weighted Proportional Fair}
  \acro{WSN}{Wireless Sensor Network}
  \acro{WWW}{World Wide Web}
  \acro{XIXO}{(Single or Multiple) Input (Single or Multiple) Output}
  \acro{ZF}{Zero-Forcing}
  \acro{ZMCSCG}{Zero Mean Circularly Symmetric Complex Gaussian}
\end{acronym}

\maketitle
\IEEEpeerreviewmaketitle

\newcommand*{\BS}[1]{\ensuremath{\text{BS}_{#1}}}
\newcommand*{\UE}[1]{\ensuremath{\text{UE}_{#1}}}

\begin{abstract}
To increase the spectral efficiency of wireless networks without requiring full-duplex capability 
of user devices, a potential solution is the recently proposed three-node full-duplex mode. To 
realize this potential, networks employing three-node full-duplex transmissions must deal with 
self-interference and user-to-user interference, which can be managed by frequency channel and 
power allocation techniques. Whereas previous works investigated either spectral efficient or fair 
mechanisms, a scheme that balances these two metrics among users is investigated in this paper. 
This balancing scheme is based on a new solution method of the multi-objective optimization 
problem to maximize the weighted sum of the per-user spectral efficiency and the minimum spectral 
efficiency among users. The mixed integer non-linear nature of this problem is dealt by Lagrangian 
duality. Based on the proposed solution approach, a low-complexity centralized algorithm is 
developed, which relies on large scale fading measurements that can be advantageously implemented 
at the base station. Numerical results indicate that the proposed algorithm increases the spectral 
efficiency and fairness among users without the need of weighting the spectral efficiency.
An important conclusion is that managing user-to-user interference by resource assignment and 
power control is crucial for ensuring spectral efficient and fair operation of full-duplex 
networks.
\end{abstract}

\section{Introduction}\label{sec:intro}

Due to recent advancements in antenna and digital baseband technologies, as well as
radio-frequency/analog interference cancellation techniques,
in-band \ac{FD} transmissions appear as a viable alternative to traditional \ac{HD} transmission
modes~\cite{Thilina2015}.
The in-band \ac{FD} transmission mode can almost double the spectral efficiency of
conventional \ac{HD} wireless transmission modes, especially in the low transmit power
domain~\cite{Thilina2015,Mairton2016b}.
However, due to the increasing demand for supporting the transmission of large data quantities
in scarce spectrum scenarios~\cite{Thilina2015,Chung2015}, and thanks to the continued advances in
\ac{SI} cancellation technologies, \ac{FD} is being considered as a technology component beyond
small cell and short range
communications~\cite{Laughlin2015,Goyal2015}.

A viable introduction of \ac{FD} technology in cellular networks consists in making the \ac{BS}
\ac{FD}-capable, while letting the \acp{UE} operate in \ac{HD} mode.
This transmission mode is termed \ac{TNFD}~\cite{Thilina2015}, in which
only one of the three nodes
(i.e., two \acp{UE} and the cellular \ac{BS})
must have \ac{FD} and \ac{SI} suppression capability.
In a \ac{TNFD} cellular network, the \ac{FD}-capable \ac{BS} transmits
to its receiving \ac{UE},
while receiving from another \ac{UE} on the same frequency channel.

\begin{figure}
\centering
\includegraphics[width=0.6\linewidth,trim=0mm 0mm 0mm 0mm,clip]{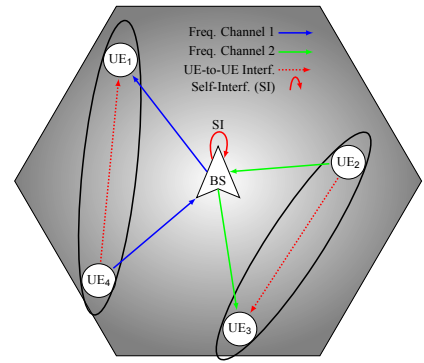}
\caption{An example of cellular network employing \ac{FD} with two \acp{UE} pairs. The \ac{BS}
selects pairs \UE{1}-\UE{4} and \UE{2}-\UE{3}, represented by the ellipses, and jointly schedules
them for \ac{FD} transmission by allocating frequency channels in the \ac{UL} and \ac{DL}.
To mitigate the UE-to-UE interference, it is advantageous to co-schedule DL/UL users for
\ac{FD} transmission that are far apart, such as \UE{1}-\UE{2} and \UE{3}-\UE{4}.}
\label{fig:scenario_tradeoff_fd}
\end{figure}

An example of a cellular network employing \ac{TNFD} with two \acp{UE} pairs is illustrated in
Figure~\ref{fig:scenario_tradeoff_fd}.
Note that apart from the inherently present \ac{SI},
\ac{FD} operation in a cellular network must also deal with the \ac{UE}-to-\ac{UE}
interference,
indicated by the red dotted lines between \UE{1}-\UE{4} and \UE{2}-\UE{3}.
The level of
\ac{UE}-to-\ac{UE} interference depends on the \acp{UE} locations and propagation environments
and their transmission powers.
To mitigate the negative effects of the interference on the spectral efficiency of the system,
coordination mechanisms are needed ~\cite{Goyal2015}.
Two key elements of such mechanisms are \ac{UE} {\it pairing} and power allocation,
that together determine which \acp{UE} are scheduled for simultaneous \ac{UL} and \ac{DL}
transmissions, and at which power \ac{UL} and \ac{DL} \acp{UE} will transmit or receive.
Consequently, it is crucial to design efficient and fair \acl{MAC} protocols and
physical layer procedures capable of supporting adequate coordination mechanisms.

A typical and natural objective for many physical layer procedures
for \ac{FD} cellular networks
proposed in the literature is to maximize
the sum spectral efficiency~\cite{Nam2015,Feng2015}.
The authors in~\cite{Nam2015} consider a joint subcarrier and power allocation problem, but
without taking into account the \ac{UE}-to-\ac{UE} interference.
The work reported in~\cite{Feng2015} considers the application of
\ac{TNFD} transmission mode in a cognitive femto-cell scenario with bidirectional transmissions
from \acp{UE}, and develops sum-rate optimal resource allocation and power control algorithms.

Another important objective is to improve the fairness and per-user \ac{QoS} of \ac{FD}
cellular networks, as emphasized in ~\cite{Mairton2016b,Thilina2015,Cheng:15,Mairton2016a}.
In our previous work, we proposed a weighted sum spectral efficiency maximization, where the
weights represent path-loss compensation and are thus related to the rate distribution and
fairness in the system~\cite{Mairton2016b}.
The results showed that \ac{FD} cellular
networks can outperform current \ac{HD} mode if appropriate \ac{SI} cancellation and pairing
schemes are employed.
A heterogeneous statistical \ac{QoS} provisioning framework, focusing on the bidirectional \ac{FD}
link case without considering the implications of \ac{TNFD} transmissions is developed
in~\cite{Cheng:15}.
The work in \cite{Thilina2015} emphasizes the importance of fairness and that it may degrade by a 
factor of two compared with \ac{HD} communications. However, the authors do not provide power 
control and channel allocation schemes that are developed with such objectives in mind.
In contrast, our previous work~\cite{Mairton2016a} formulated the maximization of the
minimum spectral efficiency problem and proposed a max-min fair
power control and channel allocation solution.

However, 
the interplay between weighted sum spectral efficiency maximization
and fairness for \ac{FD} cellular networks has not been studied.
Therefore, in this work we aim to
fill this research gap by proposing a multi-objective optimization problem to maximize
simultaneously both the
weighted sum
spectral efficiency and the minimum spectral efficiency of all users.
Such an optimization problem poses technical challenges that are markedly
different from those
investigated in our previous works~\cite{Mairton2016a,Mairton2016b}. In particular, we develop an
original and new solution approach based on the use of the scalarization technique to convert the
multi-objective into a single-objective
\ac{MINLP} problem that considers jointly user pairing and UL/DL power control.
Due to the complexity of the \ac{MINLP} problem proposed, our novel solution approach relies on
Lagrangian duality and the associated centralized algorithm based on the dual problem.
This centralized
solution is tested in a realistic system simulator that indicates that the solution is
near-optimal, and increases the sum spectral efficiency and fairness among the users.
An important feature of the proposed solution is that
there is no need to consider weights in the sum spectral efficiency as done in previous works from
the literature~\cite{Mairton2016b}.
This is advantageous, because defining the weights in a weighted sum objective
function is typically cumbersome and difficult in practice.

The numerical results also indicate that measuring and taking into account
UE-to-UE interference is crucial for both overall spectral efficiency and fairness.
When UE-to-UE interference is neglected, as in~\cite{Nam2015},
the results are approximately as good as
using random assignment and equal power allocation among all users.


\vspace{-0.2cm}

\section{System Model and Problem Formulation}
\label{sec:sys_mod}
\vspace{-0.1cm}
\subsection{System Model}\vspace{-0.1cm}

We consider a hexagonal single-cell cellular system in which the \ac{BS} is \ac{FD}
capable,
while the
\acp{UE} served by the \ac{BS} are \ac{HD} capable, as illustrated by
\figref{fig:scenario_tradeoff_fd}.
In the figure, the \ac{BS} is subject to \ac{SI}, and the \acp{UE} in the
\ac{UL}
(\UE{2} and \UE{4}) cause \ac{UE}-to-\ac{UE} interference to co-scheduled \acp{UE} in the \ac{DL},
that is to \UE{3} and \UE{1} respectively.
The number of \acp{UE} in the \ac{UL} and \ac{DL} is denoted by $I$ and $J$, respectively, which
are constrained by the total number of frequency channels in the system $F$, i.e., $I \leq F$
and $J \leq F$.
The sets of \ac{UL} and \ac{DL} users are denoted by $\mathcal{I} = \{1,\ldots,I\}$ and
$\mathcal{J} = \{1,\ldots,J\}$, respectively.

In this paper, we assume that  fading is slow and frequency flat, which is an adequate model from 
the perspective of power control in existing and forthcoming cellular 
networks~\cite{Gesbert2008,3gpp.36.300}.
Let $G_{ib}$ denote the path gain
between transmitter \ac{UE}  $i$ and the \ac{BS}, $G_{bj}$ denote the path gain between
the \ac{BS} and the receiver \ac{UE} $j$, and $G_{ij}$ denote the interfering path gain between
the \ac{UL} transmitter UE $i$ and the \ac{DL} receiver UE $j$.

The vector of transmit power levels in the \ac{UL} by \ac{UE} $i$ is denoted by $\mathbf{p^u} =
[P_1^u \ldots P_I^u] $, whereas the \ac{DL} transmit powers by the \ac{BS} is denoted by
$\mathbf{p^d} = [P_1^d\ldots P_J^d]$. We define $\beta$ as the \ac{SI} cancellation coefficient to
take into account the residual \ac{SI} that leaks to the receiver. Then, the \ac{SI} power at the
receiver of the \ac{BS} is $\beta P_j^d$ when the transmit power is $P_j^d$.

As illustrated in \figref{fig:scenario_tradeoff_fd}, the UE-to-UE interference depends
heavily on the geometry of the co-scheduled UL and DL users,
which in turn is determined by the co-scheduling or {\it pairing} of \ac{UL}
and \ac{DL} users on the available frequency channels.
Therefore, \ac{UE} pairing is a key function of the system.
To capture the pairing of \ac{UE} pairs,
we define the pairing matrix, $\mathbf{X} \in \{0,1\}^{I\times J}$, such
that
\begin{equation*}
 x_{ij} =
  \begin{cases}
   1, & \text{if the UL \UE{i} is paired with the DL \UE{j}}, \\
   0, & \text{otherwise.}
  \end{cases}
\end{equation*}

The \ac{SINR} at the \ac{BS} of transmitting user $i$ and the \ac{SINR} at the receiving user $j$
of the \ac{BS} are given by
\begin{equation}\label{eq:sinr_ul_dl}
 \gamma_{i}^u = \frac{ P_i^u G_{ib} }{ \sigma^2 + \sum_{j=1}^J x_{ij} P_j^d\beta}, \;
 \gamma_{j}^d = \frac{ P_j^d G_{bj} }{ \sigma^2 + \sum_{i=1}^I x_{ij} P_i^u G_{ij}},
\end{equation}
respectively,
where $x_{ij}$ in the denominator of $\gamma_{i}^u$
accounts for the \ac{SI} at the BS, whereas $x_{ij}$ in the denominator of $\gamma_{j}^d$
accounts for the UE-to-UE interference caused by \UE{i} to \UE{j}, and $\sigma^2$ is the noise 
power.
The achievable spectral efficiency for each user is given by the Shannon equation for the \ac{UL}
and \ac{DL} as $C_{i}^u = \log_2(1 + \gamma_{i}^u)$ and $C_{j}^d =\log_2(1 + \gamma_{j}^d)$,
respectively.

In addition to the spectral efficiency, we weight the achievable spectral efficiencies by constant
weights, which are denoted by $\alpha_i^u$ and $\alpha_j^d$, respectively.
The purpose of these weights is to allow the system designer to choose between the commonly used
sum rate maximization
and important fairness related criteria such as the well known {\it
path loss compensation} typically employed in the power control of cellular
networks~\cite{Simonsson2008}.
The weights $\alpha_i^u$ and $\alpha_j^d$ can account for sum rate maximization by setting
$\alpha_i^u =\alpha_j^d=1$, or for path loss compensation by setting $\alpha_i^u =
G_{ib}^{-1}$ and $\alpha_j^d= G_{bj}^{-1}$.
\vspace{-0.2cm}
\subsection{Problem Formulation}\label{sub:prob_form}\vspace{-0.1cm}

Our goal is to maximize both the weighted sum spectral efficiency and the minimum spectral
efficiency of all users, jointly considering the assignment of \acp{UE} in the \ac{UL} and \ac{DL}
(\textit{pairing}). This multi-objective optimization problem can be transformed to a
single-objective optimization problem through the scalarization technique~\cite[Sec.
4.7.4]{Boyd2004}. We choose $(1-\mu)$ as the weight for the weighted sum spectral efficiency and
$\mu$ for the minimum spectral efficiency, where $\mu\in[0,1]$. Specifically, we formulate the
problem as
\begin{subequations}\label{eq:weight_min_prob}
\begin{align}
\underset{\mathbf{X},\mathbf{p}^u,\mathbf{p}^d}{\text{maximize}}\quad
& \Big(1-\mu\Big)\Big(\sum\nolimits_{i=1}^I \alpha_i^u C_{i}^u + \sum\nolimits_{j=1}^J \alpha_j^d 
C_{j}^d\Big)
\nonumber\\
& + \mu\; \min_{\substack{\forall i,j}} \{C_{i}^u, C_{j}^d\}\label{eq:obj_wmax}\\
\text{subject to}\quad & P_{i}^u \leq P_{\text{max}}^u, \; \forall\; i,\label{eq:power_ul}\\
\quad& P_{j}^d \leq P_{\text{max}}^d, \; \forall\; j,\label{eq:power_dl}\\
\quad& \sum\nolimits_{i=1}^I x_{ij} \leq 1, \; \forall\; j,\label{eq:one_dl_to_ul}\\
\quad& \sum\nolimits_{j=1}^J x_{ij} \leq 1, \; \forall\; i,\label{eq:one_ul_to_dl}\\
\quad& x_{ij} \in \{0,1\}, \; \forall\; i, j.\label{eq:binary_x}
\end{align}
\end{subequations}
The optimization variables are $\mathbf{p}^u$, $\mathbf{p}^d$ and $\mathbf{X}$. Constraints
\eqref{eq:power_ul} and \eqref{eq:power_dl} limit the transmit powers, whereas constraints
\eqref{eq:one_dl_to_ul}-\eqref{eq:one_ul_to_dl} assure that only one \ac{UE} in the \ac{DL} can
share the frequency resource with a \ac{UE} in
the \ac{UL} and vice-versa. For the sake of clarity, we denote the solution to
problem~\eqref{eq:weight_min_prob} as P-OPT.


Problem \eqref{eq:weight_min_prob} belongs to the category of \ac{MINLP}, which is known
for its high complexity and computational intractability~\cite{Li2006}. To find a near-to-optimal
solution to problem \eqref{eq:weight_min_prob} we establish an original approach using the dual 
problem, as described in \secref{sec:lagr_dual}.
However, the complexity of the dual problem solution might be prohibitive in practical cellular
systems, which motivates the reformulation of the dual problem in \secref{sec:cent_sol}, whose
proposed solution in Algorithm~\ref{alg:joint_rate_fair} is denoted C-HUN.
\vspace{-0.2cm}
\section{Solution Approach Based on Lagrangian Duality}\label{sec:lagr_dual}
\vspace{-0.2cm}
\subsection{Problem Transformation}\label{sub:prob_transf}\vspace{-0.1cm}

As a first step of solving problem~\eqref{eq:weight_min_prob}, we consider the standard equivalent
hypograph~\cite[Sec. 3.1.7]{Boyd2004} form of problem~\eqref{eq:weight_min_prob}, where
the new variable $t$ and two more constraints are introduced.
Note that the hypograph simplifies the problem formulation, because it allows to use a linear
function of the variable $t$ instead of the minimum between two nonlinear functions with the other
variables.

\vspace{-0.4cm}
\begin{subequations}\label{eq:max_t_prob}
\begin{align}
& \underset{\mathbf{X},\mathbf{p}^u,\mathbf{p}^d, t}{\text{maximize}}
& & \textstyle  \Big(1-\mu\Big)\Big(\sum\nolimits_{i=1}^I \alpha_i^u C_{i}^u + 
\sum\nolimits_{j=1}^J \alpha_j^d C_{j}^d\Big) + \mu t\nonumber\\
& \text{subject to} & & C_{i}^u \geq t, \forall i, \label{eq:cap_t_ul}\\
& & & C_{j}^d \geq t, \forall j \label{eq:cap_t_dl}\\
& & & \text{Constraints~\eqref{eq:power_ul}-\eqref{eq:binary_x}}, \nonumber 
\label{eq:cap_t_old_pro}
\end{align}
\end{subequations}
where $t > 0$ is an additional variable with respect to~\eqref{eq:weight_min_prob}.
Notice that problem~\eqref{eq:max_t_prob}, similarly to problem~\eqref{eq:weight_min_prob},
is a \ac{MINLP}.
\vspace{-0.1cm}
\subsection{Solution for $\mathbf{X}$ and 
$\mathbf{p}^u,\mathbf{p}^d$}\label{sub:assign_power_sol}\vspace{-0.1cm}

From problem \eqref{eq:max_t_prob}, we form the {\it partial} Lagrangian function by
taking into account constraints \eqref{eq:cap_t_ul}-\eqref{eq:cap_t_dl} and ignoring the
integer~\eqref{eq:one_dl_to_ul}-\eqref{eq:binary_x} and power allocation
constraints~\eqref{eq:power_ul}-\eqref{eq:power_dl}.
To account for these constraints, it is assumed 
that $\mathbf{X}\in\mathcal{X}$ and $\mathbf{p}^u,\mathbf{p}^d\in\mathcal{P}$,
where $\mathcal{X}$ and $\mathcal{P}$ are sets in which the assignment constraints and
power allocation constraints are fulfilled, respectively.
The Lagrange multipliers associated with problem~\eqref{eq:max_t_prob} are
$\boldsymbol{\lambda}^u,\;\boldsymbol{\lambda}^d$, where the superscripts $u$ and $d$
denote \ac{UL} and \ac{DL}, and the vectors have dimensions of $I\times 1$ and $J\times
1$, respectively.

The partial Lagrangian is a function of the Lagrange
multipliers and the optimization variables $\mathbf{X},\mathbf{p}^u,\mathbf{p}^d$ as follows:
\vspace{-0.4cm}
\begin{align}
L(\boldsymbol{\lambda}^{u},\boldsymbol{\lambda}^{d},\mathbf{X},\mathbf{p}^{u},\mathbf{p}^{d})
&\triangleq -\Big(1-\mu\Big)\Big(\sum_{i=1}^I \alpha_i^u C_{i}^u + \sum_{j=1}^J \alpha_j^d
C_{j}^d\Big) \nonumber \\
&\hspace{-2.5cm} -\mu t +\sum\nolimits_{i=1}^I \lambda_i^u \Big( t - C_{i}^u\Big) + 
\sum\nolimits_{j=1}^J \lambda_j^d
\Big(t - C_{j}^d \Big).
\end{align}
It is useful to rewrite the partial Lagrangian function as
\begin{align}\label{eq:partial_Lagr}
L(\boldsymbol{\lambda}^{u},\boldsymbol{\lambda}^{d},\mathbf{X},\mathbf{p}^{u},\mathbf{p}^{d})
&= t\Big(\sum\nolimits_{i=1}^I \lambda_i^u + \sum\nolimits_{j=1}^J \lambda_j^d -\mu \Big) 
\nonumber \\
&\hspace{-3.3cm} - \sum_{i=1}^I \Big(\lambda_i^u + (1-\mu)\alpha_i^u \Big)C_{i}^u 
 - \sum_{j=1}^J \Big(\lambda_j^d + (1-\mu)\alpha_j^d \Big)C_{j}^d.
\end{align}

Let $g(\boldsymbol{\lambda}^{u},\boldsymbol{\lambda}^{d})$ denote the dual function
obtained by minimizing the partial Lagrangian function~\eqref{eq:partial_Lagr} with
respect to the variables
$\mathbf{X},\mathbf{p}^u,\mathbf{p}^d$.
Thus,
\begin{subequations}\label{eq:g_dual_all}
\begin{align}
\hspace*{-1.8cm}
g(\boldsymbol{\lambda}^{u},\boldsymbol{\lambda}^{d}) &= \underset{\scriptscriptstyle
\mathbf{X}\in\mathcal{X}\;\mathbf{p}^u,\mathbf{p}^d\in\mathcal{P}}{\inf} 
L(\boldsymbol{\lambda}^{u},\boldsymbol{\lambda}^{d},\mathbf{X},\mathbf{p}^{u},\mathbf{p}^{d})
\label{eq:g_dual_lagrange} 
\end{align}
\begin{numcases}{g(\boldsymbol{\lambda}^{u},\boldsymbol{\lambda}^{d})=}
\underset{\scriptscriptstyle\mathbf{X}\in\mathcal{X}\;\mathbf{p}^u,\mathbf{p}^d\in\mathcal{P}}{\inf}
\textstyle\Big[\sum_i q_i^u(\mathbf{X},\mathbf{p}^u,\mathbf{p}^d) + \nonumber\\
\sum\nolimits_j q_j^d(\mathbf{X},\mathbf{p}^u,\mathbf{p}^d)\Big],  & \hspace{-2.7cm }\text{if }
$\underset{i}{\sum} \lambda_i^u + \underset{j}{\sum} \lambda_j^d = \mu$ \nonumber\\
-\infty,  & \hspace{-4.5cm}\text{otherwise,}\label{eq:g_dual}
\end{numcases}
\end{subequations}
where it follows from equality~\eqref{eq:g_dual} that the linear function $t
\big(\sum_{i=1}^I
\lambda_i^u +\sum_{j=1}^J \lambda_j^d -\mu\big)$ is lower bounded when it is identically zero,
and
\begin{subequations}\label{eq:qi_qj_expr}
\begin{align}
q_i^u (\mathbf{X},\mathbf{p}^u,\mathbf{p}^d) &\triangleq -\Big(\lambda_i^u + (1-\mu)\alpha_i^u
\Big)C_{i}^u,\\
q_j^d (\mathbf{X},\mathbf{p}^u,\mathbf{p}^d) &\triangleq -\Big(\lambda_j^d + (1-\mu)\alpha_j^d
\Big)C_{j}^d.
\end{align}
\end{subequations}

The infimum of the dual function~\eqref{eq:g_dual} is obtained when the \ac{SINR} of the
\ac{UL}-\ac{DL} pairs is maximized. We can therefore write an initial solution for the
assignment $x_{ij}$ as follows:
\begin{align}\label{eq:x_sol}
x_{ij}^{\star} =
  \begin{cases}
   1, & \text{if } (i,j)= \underset{i,j}{\argmax}\;
   \Big(q_{i}^{u}+q_{j}^{d}\Big)
   \\
   0, & \text{otherwise,}
  \end{cases}
\end{align}
where, for simplicity, we denote an ordinary pair of \ac{UL}-\ac{DL} users as $(i,j)$.
With the assignment solution given
by~\eqref{eq:x_sol} and recalling that $x_{ij}\in\mathcal{X}$,
an \ac{UL} user can be uniquely associated
with a \ac{DL} user.
However, $x_{ij}^{\star}$ is still tied through the \acp{SINR}
in the \ac{UL} and \ac{DL}, $\gamma_{i}^u$ and $\gamma_{j}^d$, respectively.
With this, the
solution for the assignment is still complex and -- through $q_{i}^{u}$ and
$q_{j}^{d}$ -- is intertwined with the optimal power allocation.

Recall that from Eq.~\eqref{eq:g_dual} we must find the infimum of the sum between terms in
Eqs.~\eqref{eq:qi_qj_expr}.
Thus, with the initial solution for the assignment problem of finding
$\mathbf{X}$, we can now evaluate the power allocation assuming that the pairs $(i,j)$ are already
formed.
The power allocation problem is formulated as follows:
\vspace*{-0.4cm}
\begin{subequations}\label{eq:max_power_alloc}
\begin{align}
\underset{\mathbf{p}^u,\mathbf{p}^d}{\text{maximize}} \quad &\hspace{-0.3cm} \sum\nolimits_{i=1}^I
\Big(\lambda_i^u +
(1-\mu)\alpha_i^u \Big)C_{i}^u+ \nonumber\\
& + \sum\nolimits_{j=1}^J \Big(\lambda_j^d + (1-\mu)\alpha_j^d \Big)C_{j}^d\\
\text{subject to} \quad& \mathbf{p}^u,\mathbf{p}^d \in \mathcal{P},
\end{align}
\end{subequations}
where the minimization of negative sums is converted to the maximization of positive sums.
From the results of~\cite{Gesbert2008}, it follows that the optimal transmit power
allocation will have either
$P_i^u$ or $P_j^d$ equal to $P_{\text{max}}^u$ or $P_{\text{max}}^d$, given that $i$ and
$j$ share a frequency channel and form a pair.
Therefore, the optimal power allocation is
found within the corner points of $(P_i^u,P_j^d)$: $(0,P_{\text{max}}^d)$,
$(P_{\text{max}}^u,0)$ or
$(P_{\text{max}}^u,P_{\text{max}}^d)$.
If a user (either in the \ac{UL} or \ac{DL}) is not sharing the resource, i.e., assigned
to a frequency channel alone, then its transmit power is simply $P_{\text{max}}^u$ or
$P_{\text{max}}^d$.
With the assignment and power allocation solutions, we now need to find the
optimal Lagrange multipliers $\boldsymbol{\lambda}^u$ and $\boldsymbol{\lambda}^d$.
\vspace{-0.2cm}
\subsection{Dual Problem Solution}\label{sub:dual_solution}\vspace{-0.1cm}

We need to find the Lagrangian multipliers $\lambda_i^u$ and $\lambda_j^d$, which also
appear
in the objective function of problem~\eqref{eq:max_power_alloc}. Given the optimal power
allocation problem~\eqref{eq:max_power_alloc}, the dual function can be written as
\vspace{-0.2cm}
\begin{align}\label{eq:g_dual_ref}
g(\boldsymbol{\lambda}^{u},\boldsymbol{\lambda}^{d}) &= -\sum_{i=1}^I \lambda_i^u C_{i}^u
- \sum_{j=1}^J \lambda_j^d C_{j}^d - h(C_{i}^u,C_{j}^d),
\end{align}
where the term $h(C_{i}^u,C_{j}^d)$ is the weighted sum of \ac{UL} and \ac{DL} spectral
efficiencies, which are independent of $\lambda_i^u$ and $\lambda_j^d$, and do not impact
the dual problem.
Therefore, the dual problem of \eqref{eq:max_power_alloc} can be formulated as
\begin{subequations}\label{eq:dual_prob}
\begin{align}
& \underset{\boldsymbol{\lambda}^{u},\;\boldsymbol{\lambda}^{u}}{\text{minimize}} & &
\sum\nolimits_{i=1}^I \lambda_i^u C_{i}^u + \sum\nolimits_{j=1}^J \lambda_j^d C_{j}^d \\
& \text{subject to} & & \sum\nolimits_{i=1}^I \lambda_i^u + \sum\nolimits_{j=1}^J \lambda_j^d =
\mu\label{eq:const_lambda},\\
& & & \lambda_i^u,\; \lambda_j^d \geq 0, \forall i,j,
\end{align}
\end{subequations}
where the maximization of negative sums is converted to the minimization of positive sums.
The dual problem~\eqref{eq:dual_prob} is a \ac{LP} problem in the variables $\lambda_i^u$
and $\lambda_j^d$.

It is convenient to rewrite the dual problem~\eqref{eq:dual_prob} in the standard \ac{LP}
form as~\cite[Sec. 4.2]{Griva2009}:
\begin{subequations}\label{eq:dual_prob_std_lp}
\begin{align}
& \underset{\boldsymbol{\lambda}}{\text{minimize}} & &
\mathbf{c}^T \boldsymbol{\lambda} \\
& \text{subject to} & & \mathbf{a} \boldsymbol{\lambda} = b\label{eq:const_lambda_std},\\
& & & \boldsymbol{\lambda} \geq 0,
\end{align}
\end{subequations}
where $\mathbf{c} = [C_{1}^u\;\ldots\; C_{I}^u\; C_{1}^d\; \ldots\; C_{J}^d]^T$, the variable
vector is $\boldsymbol{\lambda} = [\lambda_1^u\;\ldots\; \lambda_I^u\; \lambda_1^d\; \ldots\;
\lambda_j^d ]^T$, the constraint vector $\mathbf{a} = \mathbf{1}^T$, and $b=\mu$. Since
$\mathbf{a}$ has rank 1,
we can separate the components of $\boldsymbol{\lambda}$ into two subvectors~\cite[Sec.
4.3]{Griva2009}, one consisting of $(I+J-1)$ nonbasic variables $\lambda_N$ (all of which
are zero), and another consisting of 1 basic variable $\lambda_B$, which is equal to $b$.
Therefore, we
have a single nonzero $\lambda_B$, either in the \ac{UL} or \ac{DL}, whose index $B$ corresponds
to the user with minimum spectral efficiency.

Therefore, using the results on the solution to the assignment problem \eqref{eq:x_sol}, the
optimal power allocation problem~\eqref{eq:max_power_alloc} from \secref{sub:assign_power_sol},
the dual problem~\eqref{eq:dual_prob} can be solved by checking exhaustively
which pair of \ac{UL} and \ac{DL} users jointly solve Eq.~\eqref{eq:x_sol},
where the power allocation for each pair is within the corner points of set $\mathcal{P}$.
Nevertheless, for a large number of users, this exhaustive search solution might not be practical
due to the large number of iterations.
Because of this property, we will reformulate the dual problem and propose a centralized solution
in \secref{sec:cent_sol}.
\vspace{-0.2cm}
\section{Centralized Solution based on the Lagrangian Dual 
Problem}\label{sec:cent_sol}\vspace{-0.1cm}

\subsection{Insights from the Dual Problem}\label{sub:dual_insight}\vspace{-0.1cm}
In \secref{sub:dual_solution}, we showed that the dual problem~\eqref{eq:dual_prob} maximizes the
user with minimum spectral efficiency in the system. To this end, we can initially set one
$\lambda$ equal to $\mu$ and exhaustively check which one maximizes the power allocation
problem~\eqref{eq:max_power_alloc}.
However, such exhaustive solution demands large number of iterations that depend on
the number of simultaneously served \ac{UL} and \ac{DL} users. Consequently, such solution
is not viable in practical systems.

Notice that the minimum spectral efficiency that a
user can achieve is 0, because of the binary power control solution in
problem~\eqref{eq:max_power_alloc}.
Therefore, whenever one user in the pair is not
transmitting (has zero power),
the $\lambda$ associated with that user will be nonzero, which leads to
the non-uniqueness of $\lambda$.
To reduce the complexity on the search of the
nonzero $\lambda$, we assume that for each pair there is a $\lambda$ which equals $\mu$,
whose index corresponds to the user with the
minimum spectral efficiency of that pair.
\vspace{-0.2cm}
\subsection{Centralized Solution to Reformulated Dual 
Problem}\label{sub:prof_reform}\vspace{-0.1cm}

Based on the reasoning on the non-uniqueness of $\lambda$ above and using the results from
\secref{sec:lagr_dual}, we reformulate the dual problem~\eqref{eq:dual_prob} to solve the
assignment in Eq.~\eqref{eq:x_sol}. We propose a solution that aims at jointly maximizing the sum
of the minimum
spectral efficiency of the \ac{UL}-\ac{DL} pairs and the sum spectral efficiency.
To this
end, we rewrite the solution in Eq.~\eqref{eq:x_sol} as an assignment problem given by
\begin{subequations}\label{eq:assig_prob}
\begin{align}
\underset{\mathbf{X}}{\text{maximize}} \quad & \sum\nolimits_{i=1}^I\sum\nolimits_{j=1}^J s_{ij} 
x_{ij}\\
\text{subject to} \quad& \sum\nolimits_{i=1}^I x_{ij} = 1, \; \forall j,\label{eq:dl_to_ul}\\
\quad& \sum\nolimits_{j=1}^J x_{ij} = 1, \; \forall i,\label{eq:ul_to_dl}\\
\quad& x_{ij} \in \{0,1\}, \; \forall i,j,\label{eq:binary_x_newProb}
\end{align}
\end{subequations}
where the matrix $\mathbf{S} = [s_{ijf}]\in \mathbb{R}^{I\times J}$ can be understood as
the benefit of pairing \ac{UL} user $i$ with \ac{DL} user $j$. It is given by $s_{ij}
\!= (1-\mu)(\alpha_i^u C_i^u + \alpha_j^d C_j^d) + \mu\min \{C_{i}^u, C_{j}^d\}$ for a pair
$(i,j)$ assigned to the same frequency.
Constraint~\eqref{eq:dl_to_ul} ensures that the \ac{DL} users are associated
with exactly one \ac{UL} user.
Similarly, constraint~\eqref{eq:ul_to_dl} ensures that each
\ac{UL} user must be associated with a \ac{DL} user.

Computing the optimal assignment as given by problem~\eqref{eq:assig_prob} requires
checking $(I+J)!$ assignments~\cite[Section 1]{Burkard1999}.
Alternatively, the Hungarian
algorithm can be used in a fully centralized manner~\cite[Section 3.2]{Burkard1999}, which
has worst-case complexity of $O\Big((I+J)^3\Big)$.
\algref{alg:joint_rate_fair} summarizes the steps to solve problem~\eqref{eq:assig_prob}
using the Hungarian algorithm.
The inputs to \algref{alg:joint_rate_fair} are all the path gain between \ac{UL} users,
\ac{DL} users and the \ac{BS}.
The \ac{BS} runs \algref{alg:joint_rate_fair} and
acquires or estimates the channel gains, which are measured and feedback by the served
\acp{UE} using signalling mechanisms standardized by \ac{3GPP}~\cite{3gpp.36.300}.
The most
challenging measure to obtain is the UE-to-UE interference path gain for the pair $(i,j)$, but due
to recent advances in \ac{3GPP} for \acl{D2D} communications, this measurement can be obtained
by sidelink transmissions and receptions ~\cite{3gpp.36.213}.

\begin{algorithm}
 \footnotesize
  \caption{Centralized Algorithm at the BS}\label{alg:joint_rate_fair}
 \begin{algorithmic}[1]
   \STATE Input: $\alpha_i^u,\; \alpha_i^u,\; G_{ibf},\; G_{bjf},\; G_{ijf},\; \beta,\;
   P_{\text{max}}^u,\; P_{\text{max}}^d$
   \FOR{$i=1$ \TO $I$ } 
     \FOR{$j=1$ \TO $J$ } 
        \STATE Evaluate the corner point $(P_i^u,P_j^d)$ that maximizes $s_{ij} =
             (1-\mu)(\alpha_i^u C_i^u + \alpha_j^d C_j^d) + \mu\min \{C_{i}^u, C_{j}^d\}$
             \label{alg_line:eval_cp}
     \ENDFOR
   \ENDFOR
   \STATE With $s_{ij}$, evaluate the optimal assignment using Hungarian algorithm
   \label{alg_line:eval_assign}
   \STATE \textbf{Output}: $\mathbf{X},\mathbf{p}^u,\mathbf{p}^d$\label{alg_line:output}
 \end{algorithmic}
\end{algorithm}

Once all inputs are available, the optimal power allocation for all possible pairs needs to be
evaluated (see line~\ref{alg_line:eval_cp}).
\algref{alg:joint_rate_fair} evaluates
which corner point the pair $(i,j)$
belongs to, and stores $s_{ij}$ for later use.
With $s_{ij}$ at hand, the assignment
problem~\eqref{eq:assig_prob} can be solved by
using the Hungarian algorithm~~\cite[Section 3.2]{Burkard1999} (see
line~\ref{alg_line:eval_assign}).
The outputs of the algorithm (see line~\ref{alg_line:output}) are
the assignment matrix $\mathbf{X}$, and the optimal power allocation vectors
$\mathbf{p}^u,\,\mathbf{p}^d$.
The complexity of Algorithm~\ref{alg:joint_rate_fair} hinges on creating the matrix $\mathbf{S}$,
which, recall, has a complexity $O(3IJ)$, and on the Hungarian algorithm, which has worst-case
complexity of $O\Big((I+J)^3\Big)$.

\vspace{-0.1cm}
\section{Numerical Results and Discussion}\label{sec:num_res_disc}\vspace{-0.1cm}

In this section we consider a single cell system operating in the urban micro
environment~\cite{3gpp.36.814}.
The maximum number of frequency channels is $F=25$ that corresponds to the number of
available frequency channel blocks in a 5 MHz \ac{LTE} system~\cite{3gpp.36.814}.
The total number of served \acs{UE} are $I+J=8$ and $I+J=50$, where we assume that $I=J$.
We set the weights $\alpha_i^u$ and $\alpha_j^d$ based on either sum rate maximization (SR), or
path loss compensation rule (PL). For SR, we set
$\alpha_i^u =\alpha_j^d=1$, whereas for PL we set $\alpha_i^u = G_{ib}^{-1}$ and $\alpha_j^d=
G_{bj}^{-1}$. The parameters of the simulations are set according to \tabref{tab:sim_param}.

To evaluate the performance of the proposed centralized solution in~\algref{alg:joint_rate_fair},
we use the RUdimentary Network Emulator (RUNE) as a basic platform for system simulations and
extended it to \ac{FD} cellular networks~\cite{Zander2001}.
The RUNE \ac{FD} simulation tool allows to generate the environment of \tabref{tab:sim_param}
and to perform Monte Carlo simulations using either an exhaustive search algorithm to solve
problem~\eqref{eq:weight_min_prob} or the centralized Hungarian solution.

\begin{table}
\vspace{-0.2cm}
 \caption{Simulation parameters}\label{tab:sim_param}
 \scriptsize
 \begin{tabularx}{\columnwidth}{l|l}
    \hline 
    \textbf{Parameter}                         & \textbf{Value} \\
    \hline 
    Cell radius                                & \unit{100}{m} \\
    Number of \ac{UL} \acp{UE} $[I=J]$         & $[4\; 25]$  \\
    Monte Carlo iterations                     & $400$ \\ \hline
    Carrier frequency                          & \unit{2.5}{GHz} \\
    System bandwidth                           & \unit{5}{MHz} \\
    Number of freq. channels $[F]$             & $[4\; 25]$  \\
    \ac{LOS} path-loss model                   & $34.96 + 22.7\log_{10}(d)$\\
    \ac{NLOS} path-loss model                  & $33.36 + 38.35\log_{10}(d)$\\
    Shadowing st. dev. \ac{LOS} and \ac{NLOS}  & \unit{3}{dB} and \unit{4}{dB}\\
    Thermal noise power $[\sigma^2]$           & \unit{-116.4}{dBm}/channel \\
    \ac{SI} cancelling level $[\beta]$         & $-100$~\unit{}{dB}\\ \hline
    Max power $[P^u_{\text{max}}]=[P^d_{\text{max}}]$& \unit{24}{dBm}\\
    \hline
\end{tabularx}
\end{table}

Initially, we compare the optimality gap between the exhaustive search solution of
problem~\eqref{eq:weight_min_prob}, named P-OPT, and our proposed solution using
\algref{alg:joint_rate_fair} with the optimal power allocation and the
centralized Hungarian algorithm for the assignment, named C-HUN.
In the following, we compare our proposed centralized solution with a basic \ac{FD} solution with
random assignment and equal power allocation for \ac{UL} and \ac{DL} users, named herein as R-EPA.
In addition, we also consider a modified version of C-HUN that does not take into account
\ac{UE}-to-\ac{UE} interference, named C-NINT. The motivation for C-NINT is to analyse how
important the consideration of \ac{UE}-to-\ac{UE} interference is to the fairness and
the sum spectral efficiency of the system.

\begin{figure}
\vspace{-0.2cm}
\centering
\includegraphics[width=0.7\linewidth,trim=0mm 0mm 0mm 2mm,,clip]{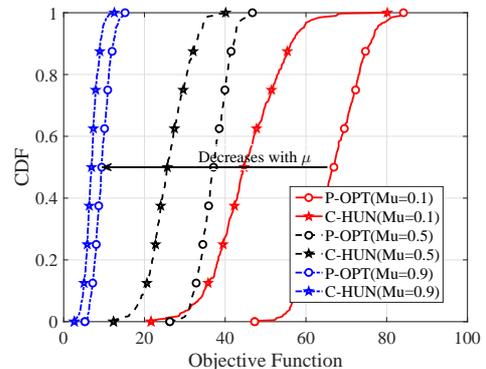}
\vspace{-0.2cm}
\caption{\ac{CDF} of the objective function in Eq.~\eqref{eq:obj_wmax} for different values of
$\mu$. We notice that the optimality gap between P-OPT and C-HUN decreases with $\mu$. 
Moreover, the objective function decreases with $\mu$, which is expected because of the reduction
of the term with sum spectral efficiency.}
\label{fig:CDF_comp_Opt_Obj_Fun}
\end{figure}

\figref{fig:CDF_comp_Opt_Obj_Fun} shows the objective function in Eq.~\eqref{eq:obj_wmax}
between P-OPT and C-HUN as a measure of the optimality gap.
We assume a small system with reduced
number of users, 4 \ac{UL} and \ac{DL} users, and frequency channels, where we assume $\mu\!=\!
[0.1\; 0.5\;0.9]$ and with $\alpha_i^u\!=\!\alpha_j^d\!=\!1$ to represent SR. Moreover, we
consider a \ac{SI} cancelling level of $\beta\!=$\unit{-100}{dB}.
Notice that the difference between the P-OPT and C-HUN decreases when $\mu$ increases.
For instance, for $\mu\!=\!0.1$ the relative difference between P-OPT and C-HUN is approximately
\unit{33}{\%}, whereas for $\mu\!=\!0.9$ this difference decreases to \unit{26}{\%}.
In addition,
the value of the objective function also decreases with $\mu$ because the term with the sum
spectral efficiency also decreases.

\begin{figure}
\vspace{-0.1cm}
\centering
\includegraphics[width=0.7\linewidth,trim=0mm 0mm 0mm
2mm,,clip]{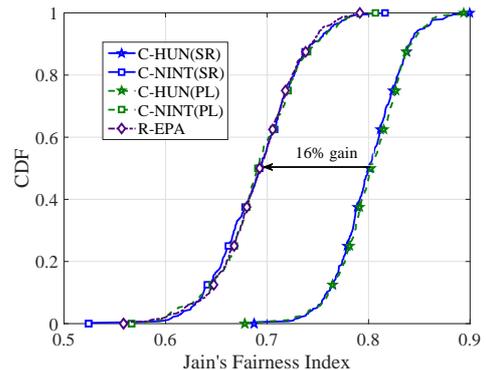}
\vspace{-0.2cm}
\caption{\ac{CDF} of Jain's fairness index for $\mu\!=0.9$ and different different weights of
$\alpha_i^u$ and $\alpha_j^d$. Notice that C-HUN achieves similar performance for SR and PL, which
implies that for high values of $\mu$ SR is enough to achieve high fairness in the system. Also,
C-NINT is as good as a R-EPA, but with higher complexity. }
\label{fig:Mu0.5_CDF_comp_Opt_Jain_Index}
\end{figure}

\figref{fig:Mu0.5_CDF_comp_Opt_Jain_Index} shows Jain's fairness index for the proposed solution
C-HUN, the modified solution C-NINT that does not consider \ac{UE}-to-\ac{UE} interference, and
the basic benchmark solution R-EPA. We assume a system fully loaded with 25 \ac{UL}, \ac{DL}
users, and frequency channels, where we analyse the impact of the solutions for different weights
of $\alpha_i^u$ and $\alpha_j^d$, which are denoted SR for sum rate maximization, and PL for
path-loss compensation. The value of $\mu$ is 0.9, which implies that we aim at a more fair
scenario. The \ac{SI} cancelling level is \unit{-100}{dB}, i.e., $\beta=$\unit{-100}{dB}.
We notice that the difference between SR and PL for the proposed solution C-HUN is negligible,
which implies that we can achieve similar levels of fairness without using weights on $\alpha_i^u$
and $\alpha_j^d$.
Conversely, there is a gain of approximately \unit{16}{\%} between C-HUN and
C-NINT at the 50-th percentile irrespectively of the weights on $\alpha_i^u$ and $\alpha_j^d$.
In addition, there is practically no difference between C-NINT and R-EPA, which implies that using
advanced solutions for pairing and power allocation without considering \ac{UE}-to-\ac{UE}
interference bring losses to the system, and is as good as doing everything randomly and setting
maximum power to all users.
Thus, our proposed solution C-HUN is able to improve fairness in the system by approximately
\unit{16}{\%} in comparison with the benchmark solution R-EPA.

\begin{figure}
\vspace{-0.2cm}
\centering
\includegraphics[width=0.7\linewidth,trim=0mm 0mm 0mm
2mm,,clip]{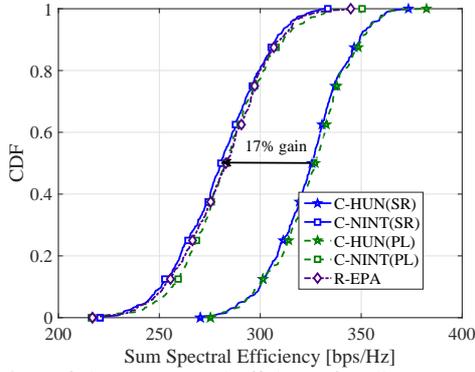}
\vspace{-0.2cm}
\caption{\ac{CDF} of the sum spectral efficiency for all users. We notice that C-HUN is also able
to improve the sum spectral efficiency with respect to C-NINT and R-EPA. Moreover, C-HUN with SR
has practically the same performance as PL, implying that the weights on $\alpha_i^u$ and
$\alpha_j^d$ are not necessary for high values of $\mu$.}
\label{fig:Mu0.5_CDF_comp_Opt_Sum_SE}
\end{figure}

\figref{fig:Mu0.5_CDF_comp_Opt_Sum_SE} shows the sum spectral efficiency of the system for C-HUN,
C-NINT, and R-EPA, where we assume the same parameters as the ones used for
\figref{fig:Mu0.5_CDF_comp_Opt_Jain_Index}.
As before, the difference between SR and PL for the proposed solution C-HUN is negligible,
implying that also for the sum spectral efficiency there is practically no difference between
using weights on $\alpha_i^u$ and $\alpha_j^d$.
As noted earlier, there is a gain of approximately
\unit{17}{\%} between C-HUN and C-NINT at the 50-th percentile for SR weights on $\alpha_i^u$ and
$\alpha_j^d$.
Also, note that C-NINT for SR is slightly outperformed by R-EPA, and for PL, C-NINT is as good as
R-EPA.
This clearly shows that
\ac{UE}-to-\ac{UE} interference needs to be taken into account if the system wants to maximize sum
spectral efficiency. Therefore, C-HUN improves the sum spectral efficiency of the system by
approximately \unit{17}{\%} in comparison with the benchmark solution R-EPA.
Overall, we notice
that when $\mu$ is high, there is no need to use weights on $\alpha_i^u$ and $\alpha_j^d$ to
improve the sum spectral efficiency or/and fairness of the system.
In addition, the
\ac{UE}-to-\ac{UE} interference needs to be taken into account if the system also wants to improve
the sum spectral efficiency or/and fairness.
Regardless of how the assignment and power allocation are performed, if \ac{UE}-to-\ac{UE}
interference is not taken into account, the results are approximately as good as using random
assignment and equal power allocation among all users.
\vspace{-0.6cm}
\section{Conclusion}\label{sec:concl}\vspace{-0.1cm}
In this paper we investigated the multi-objective problem of balancing sum spectral efficiency and
fairness among users in \ac{FD} cellular networks. Specifically, we scalarized the problem to
maximize the weighted sum
spectral efficiency and
the minimum spectral efficiency of the users, where now we can tune the weights to move towards sum
spectral efficiency maximization or fairness.
This problem was posed as a mixed integer nonlinear optimization, and given its high complexity,
we resorted to Lagrangian duality. However, the solution of the dual problem was still
prohibitive for networks with large number of users. Thus, we used the observations and results of
the dual problem to propose a low-complexity centralized solution that can be implemented at the
cellular base station.
The numerical results showed that our centralized solution improved the sum spectral efficiency
and fairness regardless of the weights on the sum spectral efficiencies of \ac{UL} and \ac{DL}
users. Furthermore, the \ac{UE}-to-\ac{UE} interference needs to
be taken into account, because otherwise irrespectively of how the assignment and power allocation
are performed, the performance in terms of sum spectral efficiency and fairness will be close to a
random assignment and equal power allocation among users.
\vspace{-0.4cm}
%
\bibliographystyle{IEEEtran}
\bibliography{FDref}

\end{document}